\documentclass[aip, pof, reprint]{revtex4-1}
\usepackage{graphicx}
\usepackage{color}
\usepackage{amssymb}
\usepackage{amsmath}
\usepackage{tabularx}
\usepackage{bm}
\usepackage{hyperref}
\usepackage{ltxtable}
\usepackage{natbib} 
\usepackage{longtable}
\usepackage{natbib}

\usepackage[usenames,dvipsnames]{xcolor}
\hypersetup{
colorlinks,
citecolor=blue,
filecolor=blue,
linkcolor=blue,
urlcolor=blue}

\newcommand{\be}{\begin{equation}}
\newcommand{\ee}{\end{equation}} 
\newcommand{\bea}{\begin{eqnarray}}
\newcommand{\eea}{\end{eqnarray}}

\begin{document}

\title{Complexity of viscous dissipation in turbulent thermal convection}

\author{Shashwat Bhattacharya}
\email{shabhatt@iitk.ac.in}
\affiliation{Department of Mechanical Engineering, Indian Institute of Technology Kanpur, Kanpur 208016, India}
\author{Ambrish Pandey}
\email{ambrish.pandey@tu-ilmenau.de}
\affiliation{Institut f\"ur Thermo- und Fluiddynamik, Technische Universit\"at Ilmenau, Ilmenau 98684, Germany }
\author{Abhishek Kumar}
\email{abhishek.kir@gmail.com}
\affiliation{Applied Mathematics Research Centre, Coventry University, Coventry, CV15FB,
The United Kingdom}
\author{Mahendra K. Verma}
\email{mkv@iitk.ac.in}
\affiliation{Department of Physics, Indian Institute of Technology Kanpur, Kanpur 208016, India}

\date{March 2018}%

\begin{abstract}
Using direct numerical simulations of turbulent thermal convection for Rayleigh number ($\mathrm{Ra}$) between $10^6$ and $10^8$ and unit Prandtl number, we derive scaling relations for viscous dissipation in the bulk and in the boundary layers.  We show that contrary to the general belief, the total viscous dissipation in the bulk is larger, albeit marginally, than that in the boundary layers. The bulk dissipation rate is similar to that in hydrodynamic turbulence with  log-normal distribution, but it differs from $(U^3/d)$ by a factor of $\mathrm{Ra}^{-0.18}$.  Viscous dissipation in the boundary layers are rarer but more intense with a stretched-exponential distribution. 
\end{abstract}
\pacs{47.55.P-, 47.27.N-, 47.27.nb}
\maketitle

Physics of hydrodynamic turbulence is quite complex, involving strong nonlinearity and  boundary effects.  To simplify, researchers have considered hydrodynamic turbulence in box away from the walls.  The turbulence in such a geometry is  statistically homogeneous and isotropic.   The physics of such  idealised flows too remain primarily unsolved, yet their energetics  is reasonably well understood.  Here, the energy supplied at large length scales cascades to intermediate scales, and then to dissipative scales~\cite{Kolmogorov:DANS1941Structure,Kolmogorov:DANS1941Dissipation}.   Thus, under steady state,  the energy supplied by the external force equals the  energy cascade rate, $\Pi_u$, and the viscous dissipation rate, $\epsilon_u$.  From dimensional analysis it has been deduced that $\epsilon_u \approx U^3/L$, where $U$ is the large-scale velocity,  $L$ is the large length scale, and the prefactor is approximately unity~\cite{McComb:book:Turbulence,Lesieur:book:Turbulence}.
 
Thermal convection is a very important problem of science and engineering.  Here too researchers have considered an idealised system called {\em Rayleigh--B\'{e}nard convection} (RBC) in which a fluid is confined between two horizontal thermal plates separated by a vertical distance of $d$; the bottom plate is hotter than the top one~\cite{Ahlers:RMP2009,Lohse:ARFM2010,Verma:NJP2017}. The kinematic viscosity ($\nu$) and thermal diffusivity ($\kappa$) are treated as constants. Additionally, the density of the fluid is considered to be a constant except for the buoyancy term of the fluid equation. The governing equations of RBC are as follows:
\bea
\partial_t \mathbf{u} + (\mathbf{u} \cdot \nabla) \mathbf{u} & = & -\nabla \sigma/\rho_0 + \alpha g \theta \hat{z} + \nu \nabla^2 \mathbf{u}, \label{eq:momentum}\\
\partial_t \theta + (\mathbf{u} \cdot \nabla) \theta & = & (\Delta/d)u_z + \kappa \nabla^2 \theta, \label{eq:energy}  \\
\nabla \cdot \mathbf{u} & = & 0, \label{eq.continuity} 
\eea  
where $\mathbf{u}$ and $\sigma$ are the velocity and pressure fields respectively,   $\theta$ is temperature fluctuation over the conduction state, $\rho_0$ and
$\alpha$ are respectively the mean density and thermal expansion coefficient of the fluid, $g$ is acceleration due to gravity, and $\Delta$ is the temperature difference between the hot and cold plates. RBC is specified by two nondimensional parameters---Rayleigh number $\mathrm{Ra} = (\alpha g \Delta d^3)/(\nu \kappa)$, which is a measure of buoyancy, and the Prandtl number $\mathrm{Pr} = \nu/\kappa$ (see supplementary material). 

For thermal convection, walls and their associated boundary layers play an important role, hence turbulence in thermal convection is more complex than hydrodynamic turbulence.  In this Letter, we focus on the  properties of the viscous dissipation in RBC.    \citet{Verzicco:JFM2003} and \citet{Zhang:JFM2017} computed the viscous dissipation rates in the bulk and in the boundary layers in RBC, and found them to be of the same order.  Here, we perform a  detailed analysis of these quantities and their probability distributions, both numerically and phenomenologically.  We will show that the walls of thermally-driven turbulence introduce interesting and generic features in the viscous dissipation.

\citet{Shraiman:PRA1990} derived an interesting exact relation that relates  the viscous dissipation rate, $\epsilon_u$, to the heat flux:
\begin{multline}
\epsilon_u = \left\langle \epsilon_u(\mathbf{r}) \right\rangle = \left\langle \frac{\nu}{2} \left( \frac{\partial u_i}{\partial x_j} + \frac{\partial u_j}{\partial x_i} \right)^2 \right\rangle  \\ = \frac{\nu^3}{d^4} \frac{\mathrm{(Nu-1)Ra}}{\mathrm{Pr}^2} =  \frac{U^3}{d}  \frac{\mathrm{(Nu-1)Ra}}{\mathrm{Re}^3 \mathrm{Pr}^2},
\label{eq:SS_exact}
\end{multline}
where $\langle \, \rangle$ denotes the volume average over the entire domain, and $u_i$ with $i = (x,y,z)$ is the $i$th the component of the velocity field. The Nusselt number, $\mathrm{Nu}$,  is the ratio of the total heat flux and the conductive heat flux, and $\mathrm{Re} = UL/\nu$ is the Reynolds number.  When the boundary layer is either absent (as in periodic box) or weak (as in the ultimate regime proposed by Kraichnan~\cite{Kraichnan:PF1962Convection}), $\mathrm{Nu} \sim (\mathrm{Ra Pr})^{1/2}$ and $\mathrm{Re} \sim (\mathrm{Ra/Pr})^{1/2}$~(See Refs.\cite{Grossmann:JFM2000,Grossmann:PRL2001,Verma:PRE2012,Verma:NJP2017}).  Substitution of these relations in Eq.~(\ref{eq:SS_exact}) yields $\epsilon_u \sim U^3/d$, similar to hydrodynamic turbulence.  In this Letter we focus on $\mathrm{Pr} \sim1$,  hence we ignore the Prandtl number dependence.

The scaling however is different for realistic RBC for which boundary layers near the plates play an important role.  Scaling arguments~\cite{Malkus:PRSA1954,Castaing:JFM1989,Grossmann:JFM2000,Grossmann:PRE2002}, experiments~\cite{Castaing:JFM1989,Qiu:PRE2002,Brown:JSM2007,Funfschilling:JFM2005,Nikolaenko:JFM2005,Ahlers:RMP2009} and numerical simulations~\cite{Verzicco:JFM2003,Stringano:JFM2006,Scheel:JFM2012, Scheel:JFM2014, Pandey:PRE2016, Pandey:PF2016} reveal that $\mathrm{Re} \sim \mathrm{Ra}^{1/2}$ and $\mathrm{Nu} \sim \mathrm{Ra}^{0.3}$, substitution of which in Eq.~(\ref{eq:SS_exact}) yields  $\epsilon_u  \ne U^3/d$, rather 
\be
\epsilon_u \sim \frac{U^3}{d}  \mathrm{Ra}^{-0.2} \sim \frac{\nu^3}{d^4}  \mathrm{Ra}^{1.3},
\label{eq:epsilon_u_total}
\ee
because $U \sim \mathrm{Re}  \sim \mathrm{Ra}^{1/2}$. This is due to the  relative suppression of the nonlinear interactions in RBC, as \citet{Pandey:PRE2016,Pandey:PF2016,Verma:NJP2017} showed that in RBC, the ratio of the nonlinear term and viscous term scales as $(UL/\nu) \mathrm{Ra}^{-0.15}$.  The aforementioned suppression of nonlinear interactions leads to weaker energy cascade $\Pi(k)$, and hence lower  viscous dissipation than the corresponding hydrodynamic turbulence.  
 
In RBC, the viscous dissipation rates in the bulk and in the boundary layers are very different.  In the following discussion, using scaling arguments and the exact relation given by Eq.~(\ref{eq:SS_exact}), we will quantify the total viscous dissipation rates in the bulk and boundary layers, $\tilde{D}_{u,\mathrm{bulk}}$ and ${\tilde{D}_{u,\mathrm{BL}}}$, as well as the corresponding average viscous dissipation rates, $\epsilon_{u,\mathrm{bulk}}$ and $\epsilon_{u,\mathrm{BL}}$, which are obtained by dividing the total dissipation rates by their respective volumes.

Grossmann and Lohse's model~\cite{Grossmann:JFM2000, Grossmann:PRL2001} assumes that $\epsilon_{u,\mathrm{bulk}} \sim U^3/d \sim \mathrm{Ra}^{3/2}$.  We find that the average viscous dissipation in the bulk scales similar to the viscous dissipation rate in the entire volume, i.e., 
\be
\epsilon_{u,\mathrm{bulk}} \sim \frac{U^3}{d}  \mathrm{Ra}^{-0.18}.
\label{eq:epsilon_u_bulk}
\ee
Since the fluid flow in the boundary layers is laminar, we expect $\epsilon_{u,\mathrm{BL}} \sim \nu U^2/\delta_u^2$, where $\delta_u$ is the thickness of the viscous boundary layer.  Hence, the ratio of the two dissipation rates is 
\bea
\frac{\epsilon_{u,\mathrm{BL}}}{\epsilon_{u,\mathrm{bulk}}} & \sim &  \mathrm{Ra}^{0.18}
\left( \frac{\nu U^2}{\delta_u^2} \right) / \left( \frac{U^3}{d} \right) \nonumber \\ 
 & \sim & \frac{1}{\mathrm{Re}} \left( \frac{d}{\delta_u} \right)^2  \mathrm{Ra}^{0.18}  \sim \left( \frac{d}{\delta_u} \right)^2  \mathrm{Ra}^{-0.32}. 
\eea
Note however that the volume of the boundary layers is much less than that of the bulk.  For simplicity, we assume that the fluid is contained in a cube of dimension $d$, then the ratio of the  volumes of the boundary layer and bulk is
\be
\frac{V_{\mathrm{BL}}}{V_{\mathrm{bulk}}}  \sim \frac{\delta_u d^2}{(d-\delta_u)^3} 
\sim \frac{\delta_u}{d},
\label{eq:volume_ratio}
\ee
because $\delta_u \ll d$ for $\mathrm{Pr} \sim 1$.  Using the above relations, we can deduce the scaling of the ratio of the total viscous dissipation rates in the boundary layer and in the bulk as
\bea
\frac{\tilde{D}_{u,\mathrm{BL}}}{\tilde{D}_{u,\mathrm{bulk}}} & \sim &   
\frac{\epsilon_{u,\mathrm{BL}}}{\epsilon_{u,\mathrm{bulk}}} \times \frac{V_{\mathrm{BL}}}{V_{\mathrm{bulk}}}
\sim  \frac{d}{\delta_u}  \mathrm{Ra}^{-0.32} . 
\label{eq:D_ratio_p}
\eea
According to Prandtl--Blassius theory~\cite{Schlichting:book:BLT2}, 
\be
\frac{\delta_u}{d} \sim \mathrm{Re}^{-1/2} \sim \mathrm{Ra}^{-1/4},
\label{eq:delta_u_PB}
\ee
which yields $\tilde{D}_{u,\mathrm{BL}}/\tilde{D}_{u,\mathrm{bulk}} \sim \mathrm{Ra}^{-0.07}$.
Thus, in RBC, the total viscous dissipation in the boundary layer and bulk are comparable to each other.  For very large $\mathrm{Ra}$, the bulk dissipation outweighs the dissipation in the boundary layer.  This is contrary to the general belief that the viscous dissipation occurs primarily in the plumes of the boundary layers.   

In this Letter, using numerical simulations we  show that $\delta_u/d$ differs slightly from Eq.~(\ref{eq:delta_u_PB}), and 
\be
\frac{\delta_u}{d} \sim \mathrm{Re}^{-0.44} \sim  (\mathrm{Ra}^{1/2})^{-0.44}  \sim \mathrm{Ra}^{-0.22},
\label{eq:delta_u_by_d}
\ee
using which we find
\be
\frac{\tilde{D}_{u,\mathrm{BL}}}{\tilde{D}_{u,\mathrm{bulk}}}  \sim \mathrm{Ra}^{-0.10}.
\label{eq:BL_Bk_ratio_theory}
\ee
Thus,
\bea
\epsilon_{u,\mathrm{BL}} & \sim & \frac{\nu U^2}{\delta_u^2} 
\sim \frac{\nu^3}{d^4} \mathrm{Ra}^{1.44},
\label{eq:epsilon_u_BL_prediction} \\
\tilde{D}_{u,\mathrm{BL}} & \sim & \epsilon_{u,\mathrm{BL}} \, \delta_u d^2 \sim  \frac{\nu^3}{d} \mathrm{Ra}^{1.22}, \label{eq:D_u_BL_prediction} \\
\tilde{D}_{u,\mathrm{bulk}} & \sim & \epsilon_{u,\mathrm{bulk}} \, d^3 \sim  \frac{\nu^3}{d} \mathrm{Ra}^{1.32}.  \label{eq:D_u_bulk_prediction}
\eea
Interestingly, $\tilde{D}_{u,\mathrm{BL}} \sim d^2 \nu U^2/\delta_u \sim (\nu^3/d)\mathrm{Ra}^{5/4}$, as assumed in Grossmann and Lohse's model~\cite{Grossmann:JFM2000, Grossmann:PRL2001}.

\begin{figure}[b]
\includegraphics[scale=0.38]{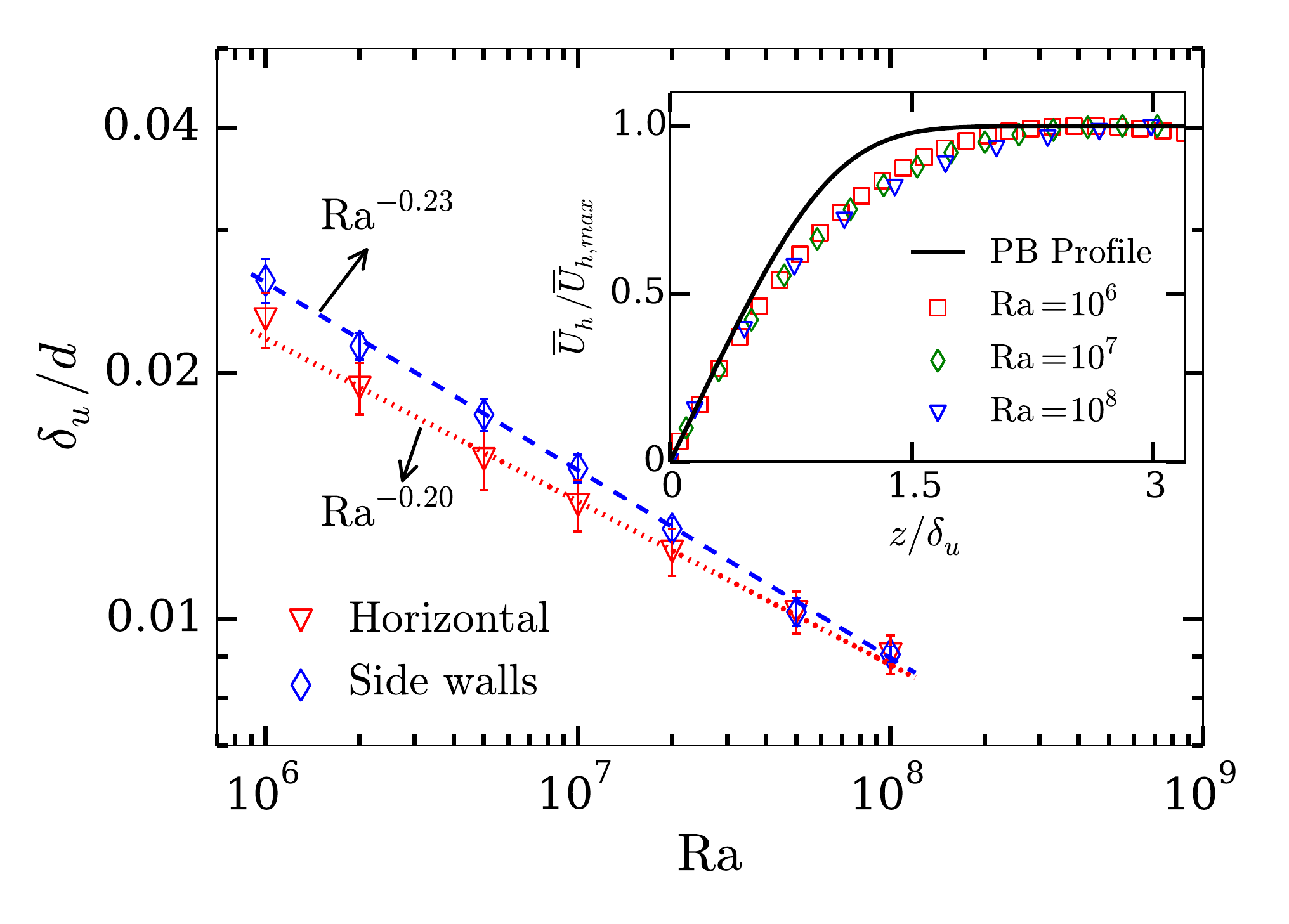}
\caption{Plot of normalized boundary layer thickness $\delta_u/d$ vs. $\mathrm{Ra}$ for horizontal and vertical plates.   Best fits are depicted as dashed and dotted lines. Inset shows the comparison of horizontal velocity profiles near the bottom plate with the Prandtl--Blasius profile (solid black line).}
\label{fig:BL_thickness}
\end{figure} 

\begin{table*}
\caption{Details of our direct numerical simulations performed in a unit box for $\mathrm{Pr}=1$: the Rayleigh Number (Ra), the kinematic viscosity ($\nu$), the Reynolds Number (Re), the ratio of the Kolmogorov length scale ($\eta$) to the average mesh width $\Delta x_{\mathrm{avg}}$, the Nusselt Number (Nu), the Nusselt number deduced from $\epsilon_u$ using Eq.~(\ref{eq:SS_exact}) ($\mathrm{Nu_S}$), number of mesh points in the viscous boundary layer ($N_{\mathrm{BL}}$), volume fraction of the boundary layer region ($V_{\mathrm{BL}}/V$), and the ratio $\tilde{D}_{u,\mathrm{BL}}/\tilde{D}_{u, \mathrm{bulk}}$.}
\begin{ruledtabular}
\begin{tabular}{c c c c c c c c c}
$\mathrm{Ra}$ & $\nu (=\kappa)$ & $\mathrm{Re}$ & $\eta/\Delta x_{\mathrm{avg}}$ & $\mathrm{Nu}$ & $\mathrm{Nu_S}$ & $N_{\mathrm{BL}}$ & $V_{\mathrm{BL}}/V$ & $\tilde{D}_{u,\mathrm{BL}}/\tilde{D}_{u, \mathrm{bulk}}$\\
\hline
$\mathrm{1 \times 10^6}$ & $0.001$ & $\mathrm{150}$ & $4.92$ & $\mathrm{8.40}$ & $8.34$ & $10$ & $0.14$ & $0.81$\\
$\mathrm{2 \times 10^6}$ & $0.0007071$ & $\mathrm{212}$ & $3.89$ & $10.1$ & $10.3$ & $8$ & $0.12$ & $0.67$\\
$\mathrm{5 \times 10^6}$ & $0.0004472$ & $\mathrm{342}$ & $2.87$ & $\mathrm{13.3}$ & $13.5$ & $7$ & $0.099$ & $0.65$\\
$\mathrm{1 \times 10^7}$ & $0.00032$ & $\mathrm{460}$ & $2.32$ & $\mathrm{16.0}$ & $15.9$ & $6$ & $0.086$ & $0.63$\\
$\mathrm{2 \times 10^7}$ & $0.0002236$ & $\mathrm{654}$ & $1.84$ & $\mathrm{20.0}$ & $20.0$ & $5$ & $0.074$ & $0.61$\\
$\mathrm{5 \times 10^7}$ & $ 0.0001414$ & $\mathrm{1080}$ & $1.36$ & $\mathrm{25.5}$ & $26.0$ & $4$ & $0.062$ & $0.57$\\
$\mathrm{1 \times 10^8}$ & $0.0001$ & $\mathrm{1540}$ & $1.09$ & $\mathrm{32.8}$ & $32.0$ & $4$ & $0.054$ & $0.56$\\
\end{tabular}
\label{table:SimDetails}
\end{ruledtabular}
\end{table*}

We perform direct numerical simulation of RBC and verify the  aforementioned scaling.  The simulations were performed using a finite volume code OpenFOAM~\cite{Jasak:CD2007} for $\mathrm{Pr}=1$ and Ra between $10^6$ and $10^8$ in a three-dimensional cube  of unit  dimension. We impose no-slip boundary condition at all the  walls, isothermal condition at the top and bottom walls, and adiabatic condition at the sidewalls (see supplementary material). Second-order Crank-Nicolson scheme is used for time-stepping. The values of $\nu$ and $\kappa$ used in the simulations are shown in Table \ref{table:SimDetails}, while keeping the temperature difference between the horizontal plates $\Delta=1$ for all the runs.  

We employ $256^3$ non-uniform grid points and solve the governing equations of RBC.  The grid is finer near the walls so as to adequately resolve the boundary layer. We ensure that minimum 4 grid points are in the boundary layer, thereby satisfying the criterion set by \citet{Grotzbach:JCP1983}. The ratio of the Kolmogorov length scale $\eta$ to the average mesh width $\Delta x_{\mathrm{avg}}$ remains greater than unity for each simulation run implying that the smallest length scales are being adequately resolved in our simulations.  We observe that the Nusselt numbers computed numerically using $\langle u_z \theta \rangle$ match quite closely with those computed  using $\epsilon_u$ and  Eq. (\ref{eq:SS_exact}).  See Table~\ref{table:SimDetails} for the comparison of these two Nusselt numbers. Also, to validate our code, we compute Nu for $\mathrm{Pr=6.8}$ fluid and verify that it matches quite well with the experimental value of Nu \cite{Zhou:JFM2012}. We further remark that our simulations capture the large-scale quantities---volume-averaged viscous dissipation and Nusselt number---quite well; such quantities are not affected significantly by discretization errors at very small scales.  Note that spectral method is more accurate but more complex than a finite volume method; yet a sufficiently-resolved finite volume code is quite appropriate for studying large-scale quantities.

\begin{figure}[htbp]
\includegraphics[scale=0.35]{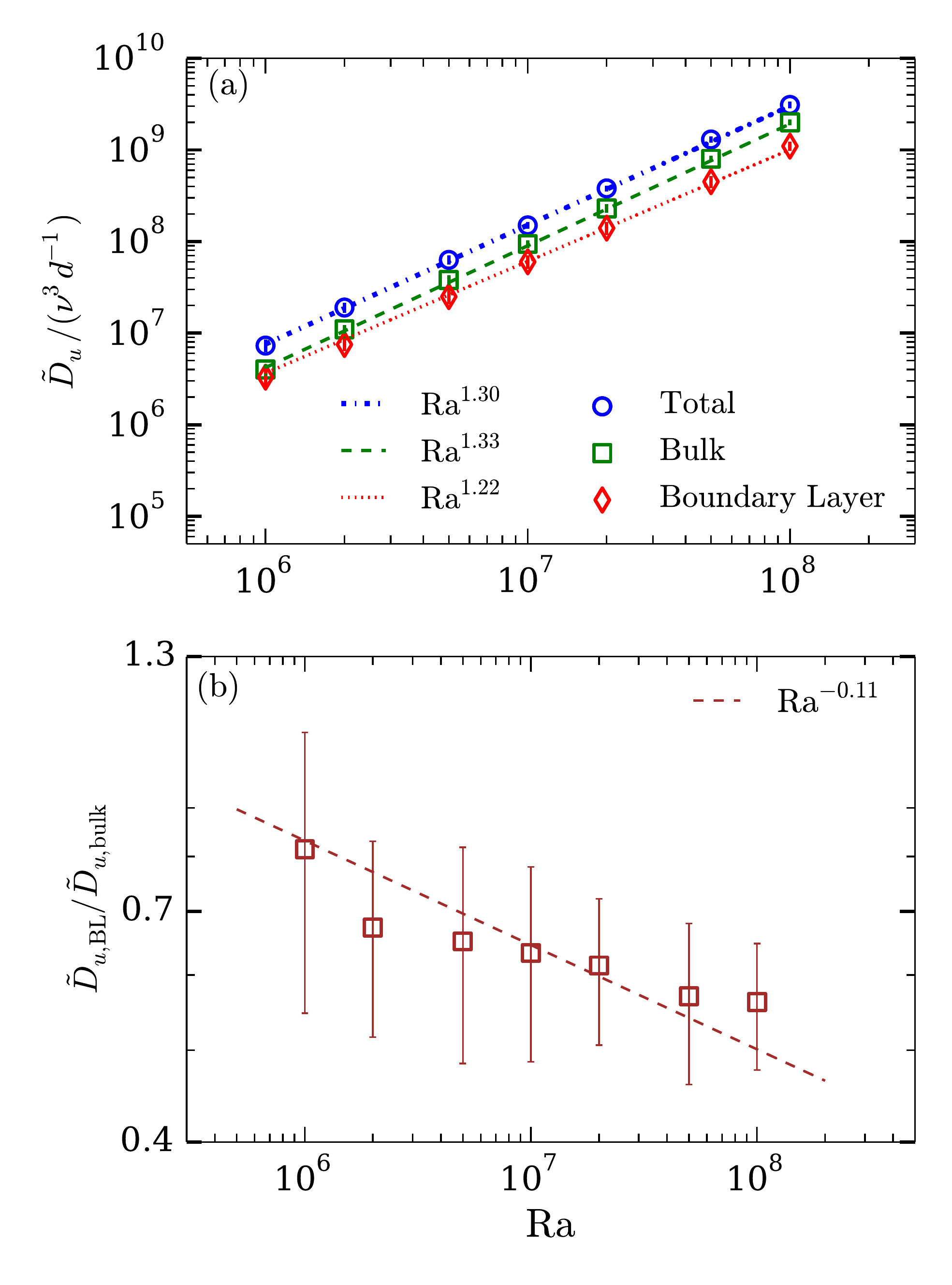}
\caption{(a) Plots of the viscous dissipation rates $\tilde{D}_u$---total,  bulk, and in the boundary layer---vs.~$\mathrm{Ra}$. (b) Plot of the dissipation rate ratio, $\tilde{D}_{u,\mathrm{BL}}/\tilde{D}_{u, \mathrm{bulk}}$, vs.~Ra that varies as $\mathrm{Ra}^{-0.11}$.  }
\label{fig:contribution}
\end{figure}

\begin{figure}[htbp]
\includegraphics[scale=1.2]{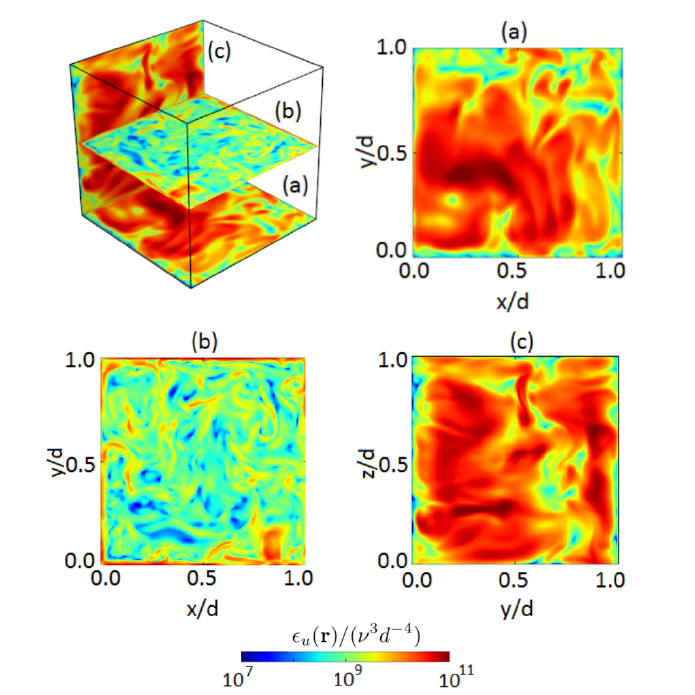}
\caption{For $\mathrm{Ra} = 10^8$: Spatial distribution of normalized viscous dissipation rate $\epsilon_u({\bf r})/(\nu^3 d^{-4})$  in planes (a) in the bottom boundary layer at $z = 2\delta_u/3$, (b) in the bulk at $z = 0.5d$, and (c) in one of the sidewall boundary layers at $x=2\delta_u/3$.}
\label{fig:diss_contour}
\end{figure}

First we compute the thickness of the boundary  layer, $\delta_u$,  for all our runs. For the same, we compute the root mean square horizontal velocity in each horizontal plane and estimate $\delta_u$ as the vertical height of the intersection of the tangent to the  profile at its local maximum with the slope of the profile at the plates~\cite{Qiu:PRE1998a, Scheel:JFM2012, Shi:JFM2012}.   Similar computations are performed for the side walls.  In Fig.~\ref{fig:BL_thickness} we plot $\delta_u$ for the horizontal and side walls.  The best fit curves of the data yield
\bea
\mbox{At thermal plates:  } &&  \delta_u/d  = 0.35  \mathrm{Ra}^{-0.20}, \\
\mbox{At sidewalls:  } &&  \delta_u/d = 0.62  \mathrm{Ra}^{-0.23}, \\
\mbox{Average: } && \delta_u/d = 0.52 \mathrm{Ra}^{-0.22}, 
\label{eq:delta_u_numerical}
\eea
with the errors in the exponents and prefactors being $\approx 0.002$ and 0.01 respectively. In Fig.~\ref{fig:BL_thickness}, we plot the horizontal and sidewall boundary layer thicknesses against Ra. These results, a key ingredient of our scaling arguments [see Eq.~(\ref{eq:delta_u_by_d})], are consistent with earlier works~\cite{Verzicco:JFM1999, Verzicco:JFM2003, Scheel:JFM2012}.    As shown in the inset of Fig.~\ref{fig:BL_thickness}, near the wall, the velocity profiles differ slightly from the Prandtl--Blasius profile, a result consistent with those of~\citet{Scheel:JFM2012} and~\citet{Shi:JFM2012}; such deviations are  attributed to the perpetual emission of plumes from the thermal boundary layers. 

We compute the ratio $V_\mathrm{BL}/V$, where $V$ is the total volume, using $\delta_u$ and Eq.~(\ref{eq:volume_ratio}).  In Table~\ref{table:SimDetails}, we list this ratio for various Ra's.  Clearly, the boundary layer occupies much less volume than the bulk, and the ratio decreases with  Ra as $\delta_u/d \propto \textrm{Ra}^{-0.22}$  [see Eq.~(\ref{eq:delta_u_by_d})]. 

After this, from the numerical data we compute the total dissipation rates in the bulk and in the boundary layer by computing $\int d\tau \epsilon_u(\mathbf{r}) $ over the respective volumes.  In Fig.~\ref{fig:contribution}(a), we plot these values for various Ra's.  Best fit curves for these data sets yield
\bea
\tilde{D}_{u, \mathrm{bulk}} & \approx & 0.05\frac{\nu^3}{d}  \mathrm{Ra}^{1.33},\label{eq:Du_Bulk_Scaling} \\
\tilde{D}_{u,\mathrm{BL}} & \approx & 0.2\frac{\nu^3}{d} \mathrm{Ra}^{1.22},
\label{eq:Du_BL_Scaling}
\eea
which are consistent with the scaling arguments presented in Eqs.~(\ref{eq:D_u_BL_prediction}, \ref{eq:D_u_bulk_prediction}). The ratio of the above quantities, plotted in Fig.~\ref{fig:contribution}(b) and listed in  Table \ref{table:SimDetails}, is
\be
\frac{\tilde{D}_{u,\mathrm{BL}}}{\tilde{D}_{u,\mathrm{bulk}}}  \approx 4\mathrm{Ra}^{-0.11},
\label{eq:Du_BL_Bk_Ratio_sim}
\ee
which is consistent with the scaling of Eq.~(\ref{eq:BL_Bk_ratio_theory}).  Note that the above ratio, listed in Table~\ref{table:SimDetails}, decreases from 0.81 to 0.56 as Ra is increased from $10^6$ to $10^8$.  Thus, bulk dissipation dominates the dissipation in the boundary layer, which is contrary to the belief that viscous dissipation primarily takes place in the boundary layer. It is however important to keep in mind that the scaling arguments take inputs from numerical simulations, such as Eq.~(\ref{eq:delta_u_numerical}) and Nusselt number scaling.

\begin{figure}[t]
\includegraphics[scale=0.38]{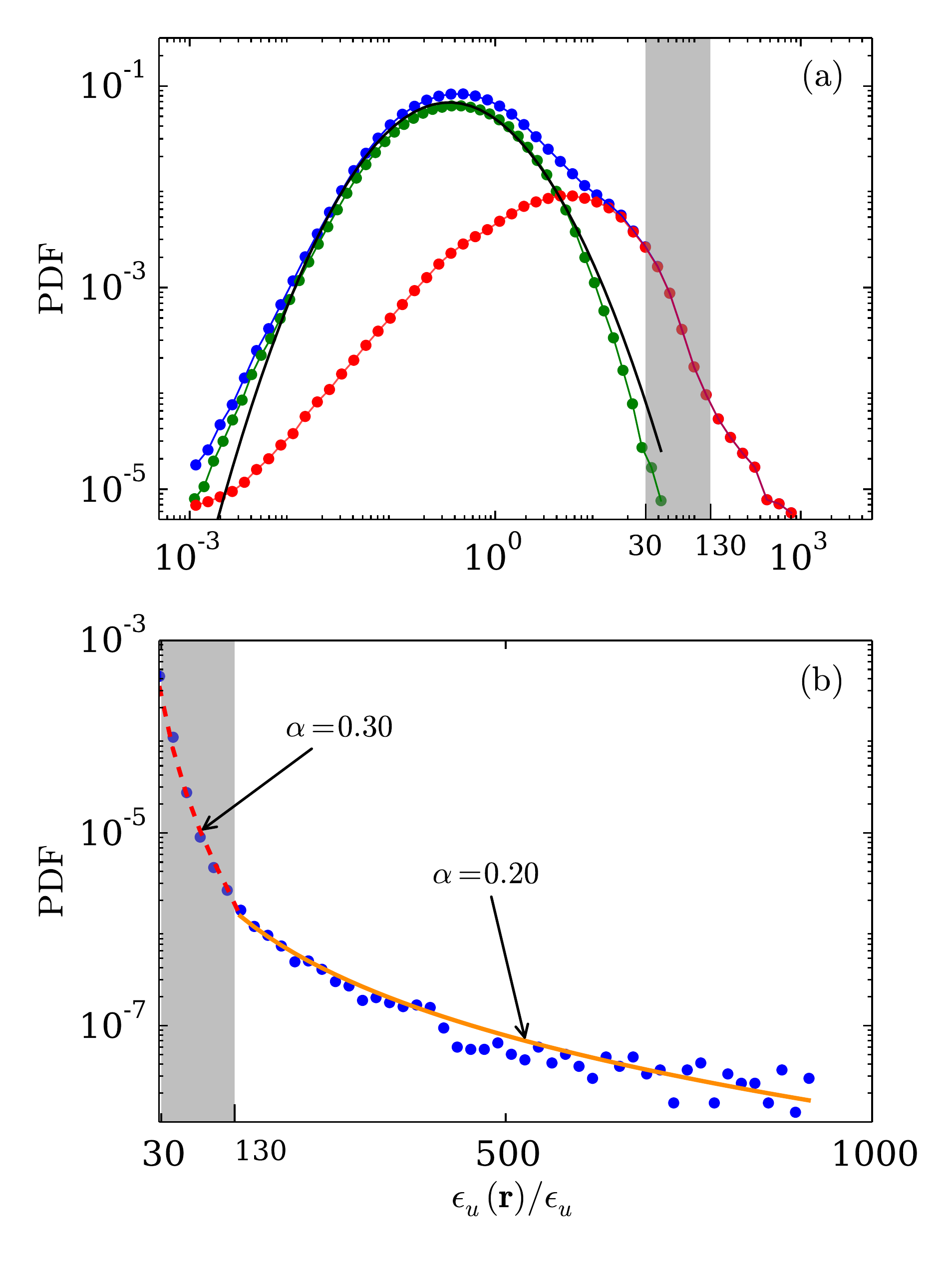}
\caption{For $\mathrm{Ra}=10^8$ and $\mathrm{Pr}=1$: (a) Probability distribution functions (PDF) of normalized local dissipation rate $\epsilon_u$  in the bulk (green), in the boundary layer (red), and in the entire volume (blue). The bulk $\epsilon_u$ has a log-normal distribution (solid black line) with $\sigma=1.2$ and $\mu=0.4$. (b) Semilog plot of the PDF of $\epsilon_u$ indicates strong tail for $\epsilon_{u,\mathrm{BL}}$ that fits well with a stretched exponential curve with $\alpha=0.30$ (dashed red line) in the shaded region, and with $\alpha=0.20$ (solid orange line) outside the region. The  shaded region is also shown in (a) for comparison.}
\label{fig:PDF}
\end{figure}
 
Thus, both scaling arguments and numerical simulations show that the bulk dissipation is weaker than that in hydrodynamic turbulence, for which $\tilde{D}_{u, \mathrm{bulk}} \sim U^3/d \sim \mathrm{Ra}^{3/2}$.  We also compute the total dissipation rate in volume $V_i = (1/4)^3 V$ located deep inside the bulk, and observe similar weak scaling with Ra (see supplementary material). Further, the viscous dissipation in the bulk dominates that in the boundary layer, albeit marginally. The boundary layer however occupies much smaller volume than the bulk.  Hence,  $\epsilon_u({\bf r})$ in the boundary layer is much more intense than in the bulk, which is illustrated in Fig.~\ref{fig:diss_contour}. Here we show density plots of normalized viscous dissipation rate $\epsilon_u({\bf r})/(\nu^3 d^{-4})$ for three planes---in the bottom and a side boundary layer, and in the bulk.  

To quantify the asymmetry of the dissipation rate  in the bulk and in the boundary layer, for $\mathrm{Ra}=10^8$, we compute the probability distribution function (PDF) of local viscous dissipation, $\epsilon_u({\bf r})$, over the full volume, the bulk, and the boundary layer. These PDFs, plotted in Fig.~\ref{fig:PDF}, reveal many important features.  Note that $\epsilon_u({\bf r}) = d\tilde{D}_u/d\tau$ with $d\tau$ as the local volume.  For $\epsilon_u({\bf r})/\epsilon_u < 20$, we observe that $\epsilon_{u,\mathrm{bulk}}({\bf r}) \gg \epsilon_{u,\mathrm{BL}}({\bf r})$, thus  average dissipation rate in the bulk is relatively weak.  But for $\epsilon_u({\bf r})/\epsilon_u > 20$, the viscous dissipation in the boundary layer dominates the bulk dissipation. 

In addition, the PDF of $\epsilon_{u,\mathrm{bulk}}$ is log-normal, similar to Obukhov's predictions~\cite{Obukhov:JGR1962} for the hydrodynamic turbulence.  See Fig.~\ref{fig:PDF}(a) for an illustration. This is consistent with the results of \citet{Kumar:PRE2014} and \citet{Verma:NJP2017}, who showed similarities between   turbulence in RBC and in hydrodynamics.    The PDF of  $\epsilon_{u,\mathrm{BL}}$ however is given by a stretched exponential---$P(\epsilon_u) \sim \beta\exp(- m\epsilon_u^{*\alpha})/\sqrt{\epsilon_u^*}$ with $\alpha \approx 0.20$ for $\epsilon_u(\mathbf{r})/\epsilon_u > 130$ and $\alpha \approx 0.30$ for $30 <\epsilon_u(\mathbf{r})/\epsilon_u < 130$ [see Fig.~\ref{fig:PDF}(b)]. Here $\epsilon^*_u$ correspond to those values of $\epsilon_u$, which are larger than the abscissa of the most probable value. This result indicates that the extreme dissipation takes place inside the boundary layer. We also carry out the PDF analysis of $\epsilon_{u,\mathrm{BL}}$ for $\mathrm{Ra}=10^6$ and $10^7$ and observe similar findings (see supplementary material). Our detailed work is consistent with earlier results  \cite{Verzicco:JFM2003,Zhang:JFM2017}.   \citet{Emran:JFM2008} reported similar PDF for the thermal dissipation rate.

We remark that by conducting a similar analysis for Pr = 6.8 and moderate Rayleigh numbers, we observe nearly identical scaling behaviour and distribution of viscous dissipation rate (see supplementary material). Thus, it can be inferred that our findings are robust.

A combination of scaling and PDF results reveals that the local viscous dissipation in the bulk, $\epsilon_{u,\mathrm{bulk}}({\bf r})$ is weak, but they add up to a significant sum due to a larger volume.  On the contrary, boundary layer exhibits extreme dissipation in a smaller volume. Interestingly, the total dissipation rate in the bulk and in the boundary layers are comparable, with bulk  dominating the boundary layer marginally.
 
Our findings clearly contrast the homogeneous-isotropic hydrodynamic turbulence and thermally-driven turbulence.  The dissipation in thermal convection has two components---$\epsilon_{u,\mathrm{bulk}}$ similar to hydrodynamic turbulence, but distinctly weaker by a factor of $\mathrm{Ra}^{-0.18}$; and $\epsilon_{u,\mathrm{BL}}$, which is unique to the flows with walls.  We believe that a similar approach could be employed to analyse the thermal dissipation rate and heat transport.

See supplementary material for a similar analysis of viscous dissipation for a larger Prandtl number $\mathrm{Pr=6.8}$ and the Rayleigh number dependence of the probability distribution function. 

\vspace{2mm}
We thank S. Fauve, R. Lakkaraju, M. Anas, and R. Samuel for useful discussions. Our numerical simulations were performed on Shaheen II at {\sc Kaust} supercomputing laboratory, Saudi Arabia, under the project k1052.  This work was supported by the research grants PLANEX/PHY/2015239 from Indian Space Research Organisation, India, and by the Department of Science and Technology, India (INT/RUS/RSF/P-03) and Russian Science Foundation Russia (RSF-16-41-02012) for the Indo-Russian project.

%

\end{document}


\title{Supplementary Material: Complexity of viscous dissipation rate in turbulent thermal convection}

\author{Shashwat Bhattacharya}
\email{shabhatt@iitk.ac.in}
\affiliation{Department of Mechanical Engineering, Indian Institute of Technology Kanpur, Kanpur 208016, India}
\author{Ambrish Pandey}
\email{ambrish.pandey@tu-ilmenau.de}
\affiliation{Institut f\"ur Thermo- und Fluiddynamik, Technische Universit\"at Ilemnau, Ilemnau 98684, Germany }
\author{Abhishek Kumar}
\email{abhishek.kir@gmail.com}
\affiliation{Applied Mathematics Research Centre, Coventry University, Coventry, CV15FB,
The United Kingdom}
\author{Mahendra K. Verma}
\email{mkv@iitk.ac.in}
\affiliation{Department of Physics, Indian Institute of Technology Kanpur, Kanpur 208016, India}

\maketitle
\section{Non-dimensionalization of the Governing Equations}
In the Letter, we described the governing equations of Rayleigh-B\'enard Convection. We nondimensionalize the governing equations by choosing $d$ as the length scale, $\sqrt{\alpha g \Delta d}$ as the velocity scale, $\Delta$ as the temperature scale, and $d/\sqrt{\alpha g \Delta d}$ as the time scale. The resulting nondimensional equations are:
\bea
\partial_t \mathbf{u} + \mathbf{u}.\nabla \mathbf{u} &=& -\nabla \sigma + \theta \hat{z} +  \sqrt{\mathrm{\frac{Pr}{Ra}}}\nabla^2 \mathbf{u}, \\
\partial_t \mathbf{\theta} + \mathbf{u}.\nabla \theta &=& u_z + \frac{1}{\sqrt{\mathrm{Ra Pr}}}\nabla^2 \theta, \\
\nabla.\mathbf{u} &=& 0,
\eea 
where $\mathrm{Ra}=\alpha g \Delta d^3/(\nu \kappa)$ is the Rayleigh number and $\mathrm{Pr}=\nu/\kappa$ is the Prandtl number. The Rayleigh and Prandtl numbers are the main governing parameters parameters of RBC.
\section{Simulation geometry, and regions of bulk and boundary layers }
\begin{figure}
\includegraphics[scale=0.18]{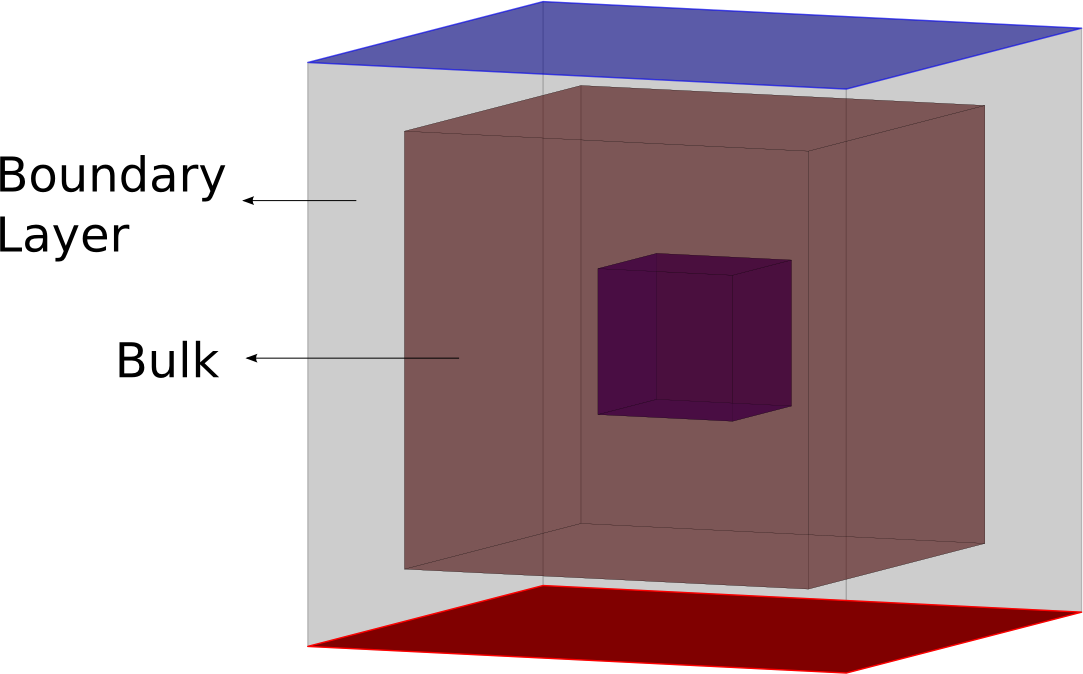}
\caption{Schematic of a cubical RBC cell with no-slip boundaries depicting the bulk (brown) and the boundary layer (gray) regions. Also shown is a cubical subvolume (purple) of length $0.25 d$ deep inside the cube.}
\label{fig:Domain}
\end{figure}

We simulate RBC in a cube with no-slip walls on all sides.  In Fig.~\ref{fig:Domain} we illustrate the box,  the bulk region, and the boundary-layers.    We compute the viscous dissipation rates in the bulk and in the boundary layers.  We also compute the viscous dissipation rate inside the innermost cube $V_i$, which is $(1/4)^3$ of the cube.
\begin{figure}
\includegraphics[scale=0.38]{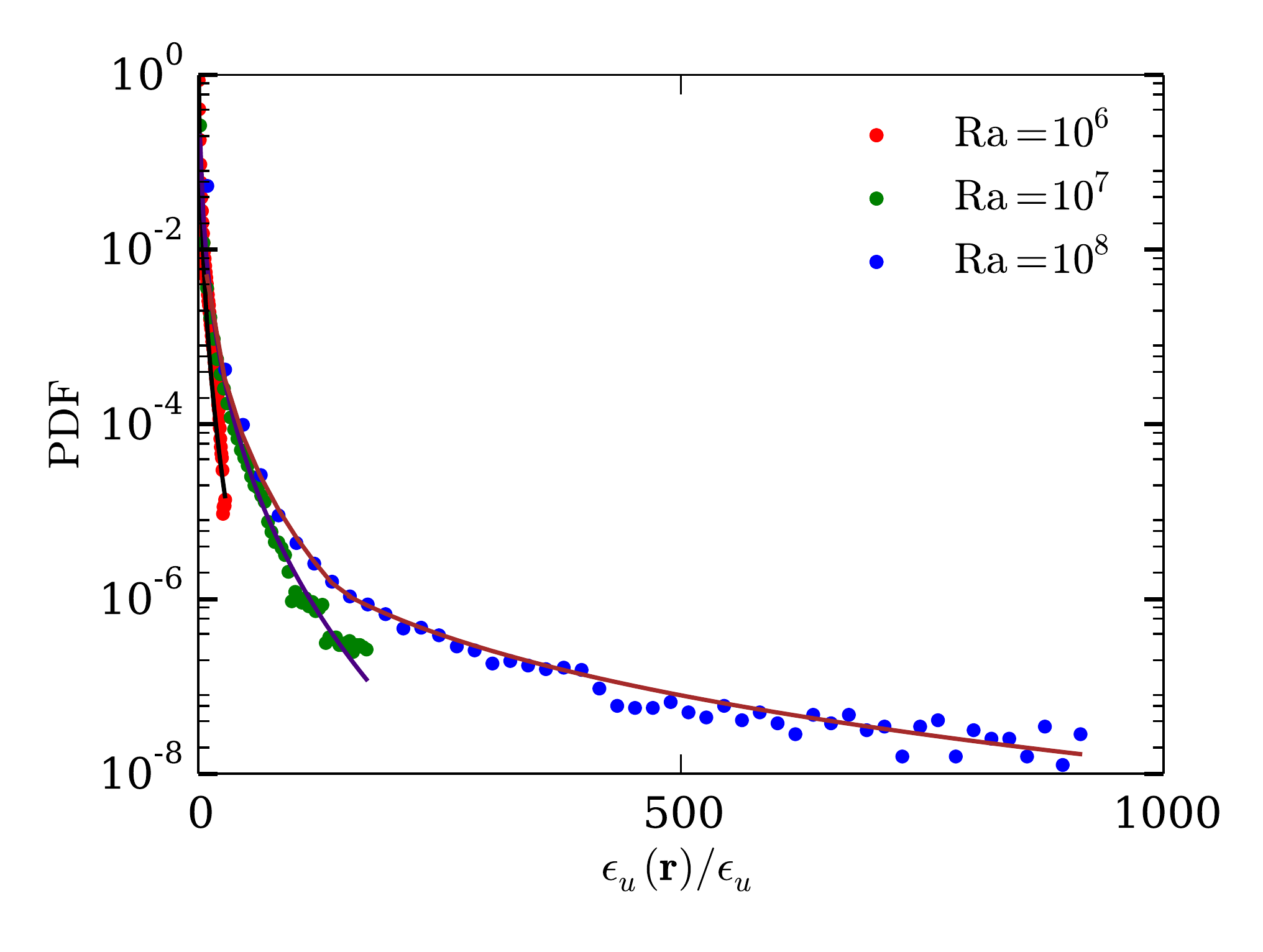}
\caption{PDFs of viscous dissipation rates for $\mathrm{Ra}=10^6$, $10^7$,  $10^8$ and $\mathrm{Pr}=1$. Tails exhibit a stretched exponential behaviour. The  brown, indigo and black curves represent the fits for $\mathrm{Ra}=10^8$, $\mathrm{Ra}=10^7$ and $\mathrm{Ra}=10^6$ respectively.}
\label{fig:PDF}
\end{figure}
\begin{figure}
\includegraphics[scale=0.40]{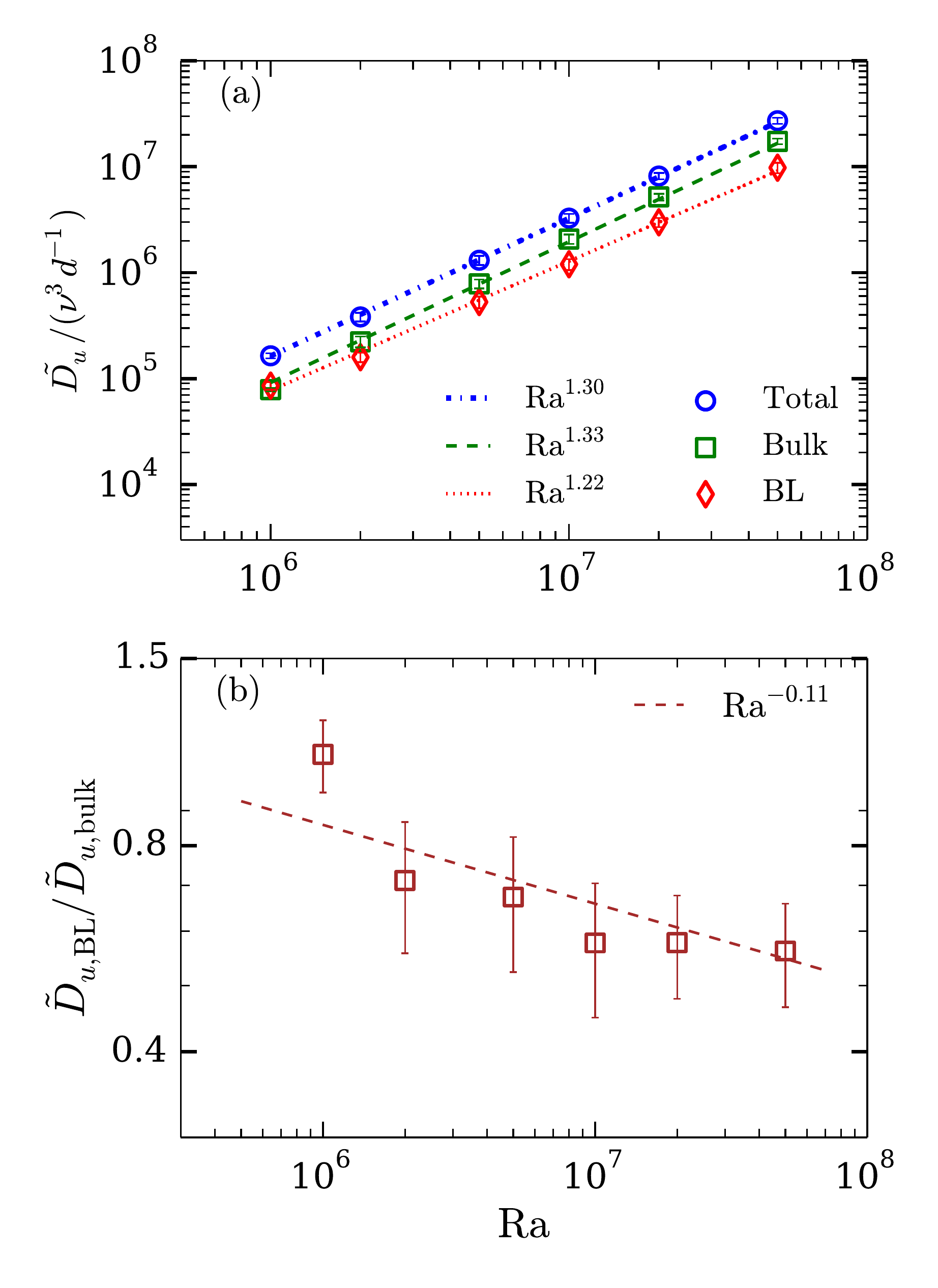}
\caption{(a) Plots of the viscous dissipation rates $\tilde{D}_u$---total,  bulk, and in the boundary layer---vs.~$\mathrm{Ra}$. (b) Plot of the dissipation rate ratio, $\tilde{D}_{u,\mathrm{BL}}/\tilde{D}_{u, \mathrm{bulk}}$, vs.~Ra that varies as $\mathrm{Ra}^{-0.11}$.}
\label{fig:contribution}
\end{figure}
\begin{figure}
\includegraphics[scale=0.40]{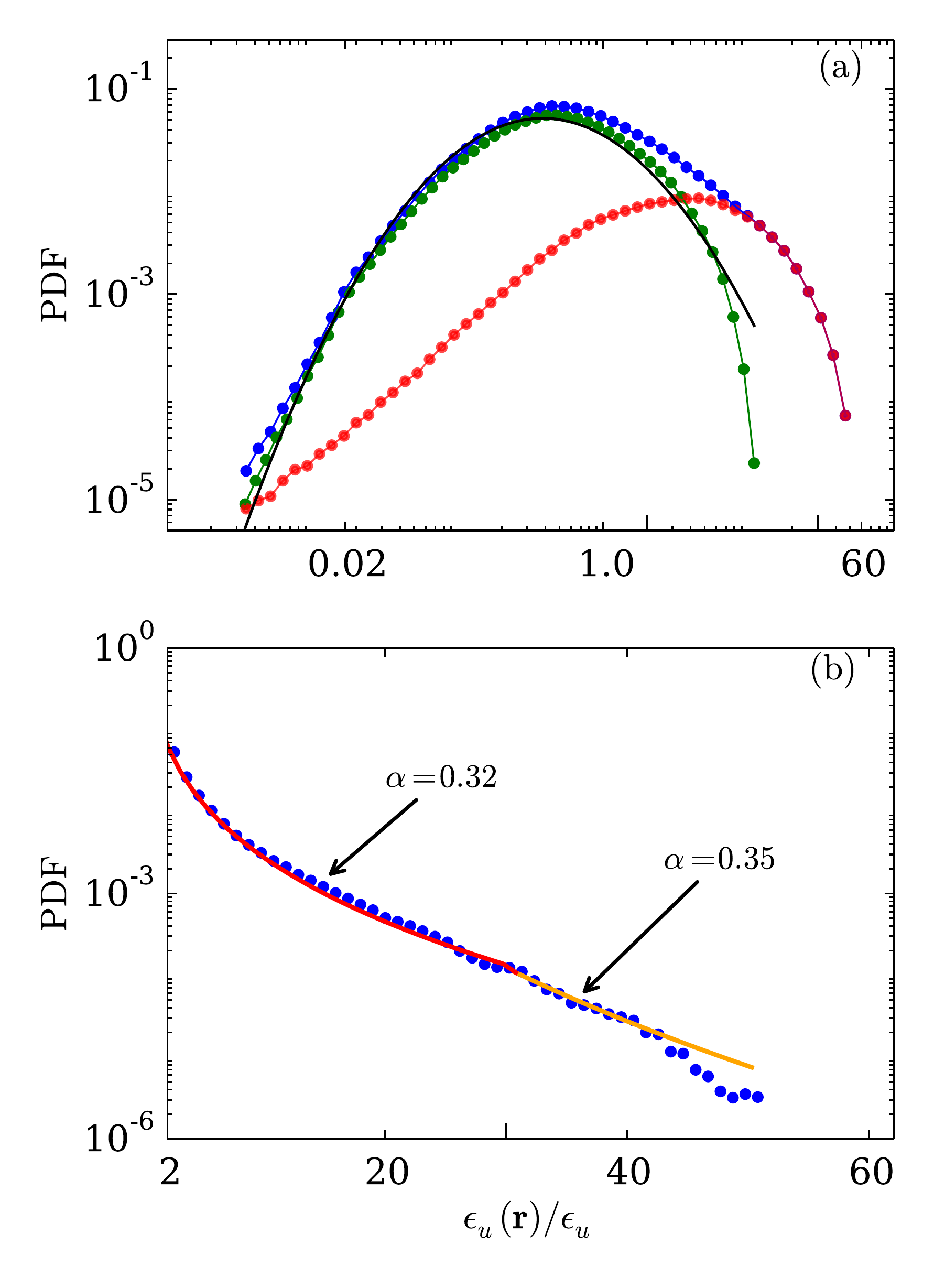}
\caption{For $\mathrm{Ra}=5 \times 10^7$ and $\mathrm{Pr}=6.8$: (a) PDF of local dissipation rates  in the bulk (green), in the boundary layer (red), and in the entire volume (blue).  The bulk $\epsilon_u$ has a log-normal distribution (solid black line). (b) Semilog plot of the PDF of $\epsilon_u$ indicates strong tail for $\epsilon_{u,\mathrm{BL}}$ that fits well with a stretched exponential curve with $\alpha=0.32$ (red line) - 0.35 (orange line).}
\label{fig:PDF_6p8}
\end{figure} 

The scaling of $\tilde{D}_u$ in these regions are as follows:
\bea
\tilde{D}_{u,\mathrm{BL}} = 0.2 \frac{\nu^3}{d} \mathrm{Ra}^{1.22}, \\
\tilde{D}_{u,\mathrm{bulk}} = 0.05 \frac{\nu^3}{d} \mathrm{Ra}^{1.33}, \\
\tilde{D}_{u,V_i} = 0.002 \frac{\nu^3}{d} \mathrm{Ra}^{1.25}. 
\eea

\section{Rayleigh-number-dependence of Probability Distribution Functions (PDF) of viscous dissipation}
In the Letter, we discussed the PDF of viscous dissipation rate, $\epsilon_u$, for $\mathrm{Ra}=10^8$.  In this section we briefly describe the PDF of $\epsilon_u$ for various Ra's in the turbulent regime. As shown in Fig.~\ref{fig:PDF}, the tail of the PDFs for all the three Rayleigh numbers exhibit stretched-exponential behavior, i.e.,
\be
P(\epsilon_u) \sim   \frac{\beta}{\sqrt{\epsilon_u^*}} \exp(-m\epsilon_u^{*\alpha}),
\ee
with $\alpha = 0.20 - 0.30$, 0.32 and 0.38 for $\mathrm{Ra}=10^8$, $10^7$ and $10^6$ respectively.  Clearly the tails are stretched more for larger Ra's.  This is expected because we expect stronger dissipation for larger Ra.

\section{Viscous dissipation for $\mathbf{{Pr}=6.8}$}

The Letter contains the description of viscous dissipation for $\mathrm{Pr}=1$ and $\mathrm{Ra}=10^6$ to $10^8$.  To show that the results described in the letter are generic, we compute the dissipation rates in the bulk and boundary layer for $\mathrm{Pr}=6.8$ and $\mathrm{Ra}=10^6$ to $5 \times 10^7$. We observe that
\bea
\tilde{D}_{u,\mathrm{bulk}} & \approx & 0.001 \frac{\nu^3}{d}\mathrm{Ra}^{1.33} \label{eq:D_bulk}, \\
\tilde{D}_{u,\mathrm{BL}} & \approx & 0.004 \frac{\nu^3}{d} \mathrm{Ra}^{1.22} \label{eq:D_BL}, \\
\frac{\tilde{D}_{u,\mathrm{BL}}}{\tilde{D}_{u,\mathrm{bulk}}} & \approx & 4 \mathrm{Ra}^{-0.11}.
\label{eq:D_ratio}
\eea \vspace{3mm}
  
Clearly, the exponents for $\mathrm{Pr}=6.8$ are very close to those for $\mathrm{Pr}=1$, thus showing that the results of the Letter are generic [see Figs.~\ref{fig:contribution}(a) and (b)]. The prefactors for $\tilde{D}_{u,\mathrm{bulk}}$ and $\tilde{D}_{u,\mathrm{BL}}$ are an order of magnitude lower than those for  $\mathrm{Pr}=1$ case, which is due to lower nonlinearity of energy flux for $\mathrm{Pr}=6.8$.

We also compute the PDF of $\epsilon_u(\mathbf{r})$   for $\mathrm{Pr}=6.8$ and $\mathrm{Ra}=5\times 10^7$ case.  The behaviour for $\mathrm{Pr}=6.8$ is very similar to that for  $\mathrm{Pr}=1$ case,  where the PDF for the bulk dissipation rate exhibits log-normal behaviour, and the PDF for the dissipation rate in the boundary layer is stretched-exponential [see Figs.~\ref{fig:PDF_6p8}(a) and (b)].  The exponent $\alpha$ for $\mathrm{Ra}=5\times 10^7$ is  larger than that for $\mathrm{Ra}=1 \times 10^8$ ($\mathrm{Pr}=1$) indicating that the latter has longer tail in the PDF.